\documentclass[journal=jacsat,manuscript=article]{achemso}

\usepackage[version=3]{mhchem} 
\usepackage{braket}
\usepackage{physics}
\usepackage{amsmath}
\usepackage{amssymb}
\usepackage[scr=rsfs]{mathalpha}
\usepackage[normalem]{ulem}
\usepackage[colorlinks]{hyperref}

\author{Laiz R. Ventura}
\affiliation[ITA]{Departamento de F\'{\i}sica, Instituto Tecnológico de Aeronáutica, 12228-900, São Jos\'{e} dos Campos, Brazil}

\author{Ramon S. da Silva}
\affiliation[ICEx]{Departmento de Física, Instituto de Ciências Exatas - ICEx, Universidade Federal Fluminense, Campus do Aterrado, Volta Redonda, RJ, 27213-45, Brazil}
\alsoaffiliation{Departamento de Física, Universidade Federal de Juiz de Fora, Juiz de Fora, MG, Brazil}
\author{Sergio R. Souza}
\affiliation[ICEx]{Departmento de Física, Instituto de Ciências Exatas - ICEx, Universidade Federal Fluminense, Campus do Aterrado, Volta Redonda, RJ, 27213-45, Brazil}
\alsoaffiliation[Second University]
{Instituto de Física, Universidade Federal do Rio de Janeiro, Rio de Janeiro, RJ, 21941-909, Brazil}
\author{Jayr Amorim}
\affiliation[ICEx]{Departmento de Física, Instituto de Ciências Exatas - ICEx, Universidade Federal Fluminense, Campus do Aterrado, Volta Redonda, RJ, 27213-45, Brazil}
\alsoaffiliation[Second University]
{Departamento de F\'{\i}sica, Instituto Tecnológico de Aeronáutica, 12228-900, São Jos\'{e} dos Campos, Brazil}
\author{Carlos E. Fellows}
\affiliation[ICEx]{Departmento de Física, Instituto de Ciências Exatas - ICEx, Universidade Federal Fluminense, Campus do Aterrado, Volta Redonda, RJ, 27213-45, Brazil}
\altaffiliation{Corresponding author}
\email{cefellows@id.uff.br}

\title[An \textsf{achemso} demo]
  {Energy shifts in predissociating levels of diatomic molecules: The case of  N$_2$ (C$''^5\Pi _u$) and N$_2$(1$^7\Sigma ^+_u$) interacting states}

\abbreviations{IR,NMR,UV}

\begin{document}

\newpage

\begin{abstract}

This work presents a perturbative calculation methodology for evaluating the
energy shifts and broadening of vibrational energy levels, caused by interactions between
bound and unbound dissociative electronic states. The method is validated against
previously semiclassical analyzed cases, demonstrating remarkable consistency. We successfully applied this approach to the N$_2$ molecule, which exhibits a strong spin-orbit
interaction between the bound C$''^5\Pi_u$  and the repulsive 1$^7\Sigma^+_u$ electronic states, around 36 cm$^{-1}$. This interaction constitutes an major pathway for N($^{2}$D) production, important in both excitation and quenching in plasma afterglows. As a result, the maximum absolute shift of 0.15 cm$^{-1}$ was found for the C$''^5\Pi_u$ ($v$ = 7) and maximum broadening of  0.45 cm$^{-1}$ was calculated for $v$ = 8, demonstrating significant perturbation of the C$''^5\Pi_u$ by the 1$^7\Sigma^+_u$ state. The results obtained were compared with  direct calculations of the predissociation rates of the C$''^5\Pi_u$ bound state, showing very good agreement.

\end{abstract}
\newpage

\section{Introduction}
\label{intro}

Perturbations between discrete vibrational levels belonging to two bound electronic states occur relatively frequently~\cite{lefebvre2004spectra,lefebvre2012perturbations}. However, the overlapping energy levels of a molecule with a dissociative continuum give rise to a pre-dissociation phenomenon, which is less frequently observed~\cite{wheeler1998predissociation,buijsse1997measurement,antonova2000predissociation,mitev2025exomol}. Herzberg \cite{herzberg2013molecular} previously classified these effects into three categories, with type I relating bound and unbound states, being the most significant in the case of diatomic molecules. As observed theoretically by Fano \cite{fano1961effects} and experimentally by  Barrow {\emph{et al.}}~\cite{barrow1966b}, an isolated discrete level that interacts with an unbound (dissociative) state undergoes a broadening and a shift in energy. An accurate estimate of the width of the broadening can be obtained once the wave functions of the discrete level and the resonant continuous level are determined, as demonstrated by Murrell and Taylor \cite{murrell1969predissociation}, and by Czarny, Felenbok and Lefebvre-Brion \cite{czarny1971high}. The direct numerical estimate of the shift has been performed in a few cases ~\cite{atabek1972evaluation,czarny1971high,julienne1975predissociation,atabek1977close} albeit in an approximate way, using model potentials to describe the interacting states, as carried out by Atabek and Lefebvre~\cite{atabek1972evaluation} or through ab-initio calculations, such as the work conducted by Julienne and Krauss \cite{julienne1975predissociation}. Other methods, such as semi-classical approximations, have been used, as in the work of Child \cite{child1975level} and Child and Lefebvre~\cite{child1977shifts}. However, solving the coupled equations is computationally intensive while the solution using semi-classical methods is complex and non-trivial.

In this work, we present an application of the perturbative solution proposed by Fano~\cite{fano1961effects}, applied to the interaction between the bound  N$_2$ (C$''^5\Pi _u$) state 
and the repulsive N$_2$ (1$^7\Sigma ^+_u$). The spin-orbit coupling (SOC) between these states is relatively strong, making them attractive for use in kinetic models. According to Wu \textit{et al.}~\cite{wu2024modeling}, this interaction plays a crucial role in both electronic excitation and quenching processes, exhibiting rate coefficients comparable to those of the N($^{4}$S) + N$_{2}$ $\rightarrow$ N($^{2}$D) + N$_{2}$ reaction. 
The results presented in the following sections demonstrate promising potential for modeling these bound-unbound state interactions, suggesting new opportunities for investigating similar interacting level systems.

The choice of the N$_2$ molecule and the electronic states C$''^5\Pi_u$ and 1$^7\Sigma^+_u$ is a natural consequence of the work recently carried out by Ventura {\emph{et al.}}~\cite{VENTURA2025125986}, where it was observed that their interaction significantly modifies the lifetime of the C$''^5\Pi_u$ state.

\section{Energy levels shift evaluations}
\label{shifts}

The unperturbed discrete energy spectrum $E_v$ of the Hamiltonian $H$, describing the vibrational states of the diatomic molecule, is obtained by solving the Schrodinger equation considering a potential $V_b$, such that

\begin{equation}
\bra{\psi_{v'}}H\ket{\psi_v}=E_v\delta_{v,v'}\;,
\label{eq:Ev}
\end{equation}

\noindent
where $\ket{\psi_v}$ is the wave function associated with the $v$-th vibrational state and $\delta_{v,v'}$ represents the Kronecker delta function.
The shift in the energy $E_v$ is given by:

\begin{equation}
\label{eq:Eshift}
    E_v' = E_v + \Delta (E_v)\;,
\end{equation}

\noindent
due to the coupling with continuum states $\ket{\phi _E}$ of energy $E$ of the Hamiltonian,
is calculated through the perturbative treatment developed by Fano~\cite{fano1961effects}, from which one writes:

\begin{equation}
\Delta (E_v) = \mathscr{P}\int dE' \frac{|V_{E'}^v|^2}{E_v-E'}\;,
\label{eq:DeltaE}
\end{equation}

\noindent
where $\mathscr{P}$ indicates ``the principal part of" and $V_E^v$  = $\bra{\phi_E}H\ket{\psi_v}$ is the potential coupling between bound and continuum states.
The latter are calculated using a repulsive potential, $V_u$, and are normalized such that

\begin{equation}
\bra{\phi_{E'}}H\ket{\phi_E}=E\delta(E-E')\;.
\label{eq:E}
\end{equation}

\noindent The Dirac delta function $\delta(E-E')$ ensures that $\ket{\phi_E}$ is normalized per unit energy, so that $|V_{E'}^v|^2$ has the dimensionality of $E$.

As a first step, in order to verify the accuracy of our calculations, a comparison with the data reported in the literature is performed.
To do this, a simple model already proposed by Child and
Lefebvre~\cite{child1977shifts} is used.
In it, the attractive potential $V_b$ is given by a Morse potential

\begin{equation}
    \label{eq:Vuc}
    V_b(r) = D[1-\exp(-\beta(r-r_e))]^2
\end{equation}

\noindent and the repulsive potential $V_u$ is represented by:

\begin{equation}
\label{eq:Vbc}
    V_u(r)=A\exp[-\alpha(r-r_c)]+B\;.
\end{equation}

\begin{figure}
	\centering
	\includegraphics[width=.9\textwidth]{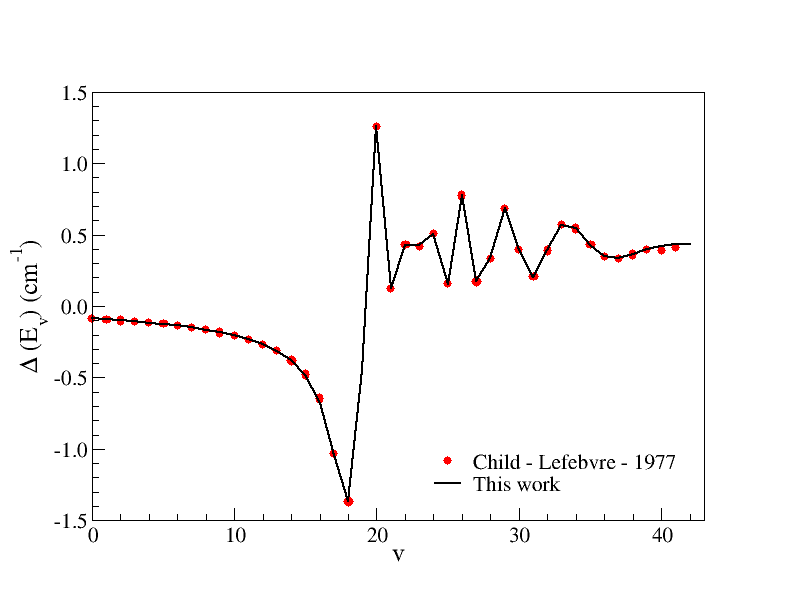}
	\caption{Comparison between the results obtained by Child and Lefebvre~\cite{child1977shifts} and those obtained in the present article.}
	\label{FIG:1}
\end{figure}

\noindent
The set of parameters used by Child and Lefebvre~\cite{child1977shifts} in their calculations,  reads: $A = 18154.95$ cm$^{-1}$, $D = 15000$ cm$^{-1}$, $B = -8000$ cm$^{-1}$, $r_e = 1.6$ \AA, $r_C = 248$ \AA, $\alpha = 2.2039$ \AA$^{-1}$ and $\beta = 1.9685$ \AA$^{-1}$. The electronic interaction, associated with $V_E^v$, is set to $\beta = 100$ cm$^{-1}$ and the value of the reduced mass $\mu$ is 8 amu.
The potential energy curves are defined so that the repulsive and attractive ones cross each other at $r_C$, which is the turning point to the right of the $v$ = 18 level of the Morse potential.
Our results are shown in Figure \ref{FIG:1}, along with those obtained by Child and
Lefebvre~\cite{child1977shifts} who solved the corresponding coupled Schrödinger equations. As can be seen, the results are in good agreement.
Figure \ref{FIG:1} illustrates excellent concordance between the results obtained from both  methodologies.

\section{The N$_2$ C$''^5\Pi _u$ and 1$^7\Sigma ^+_u$ interaction case}
\label{energy}

\begin{figure}
	\centering
	\includegraphics[width=.9\textwidth]{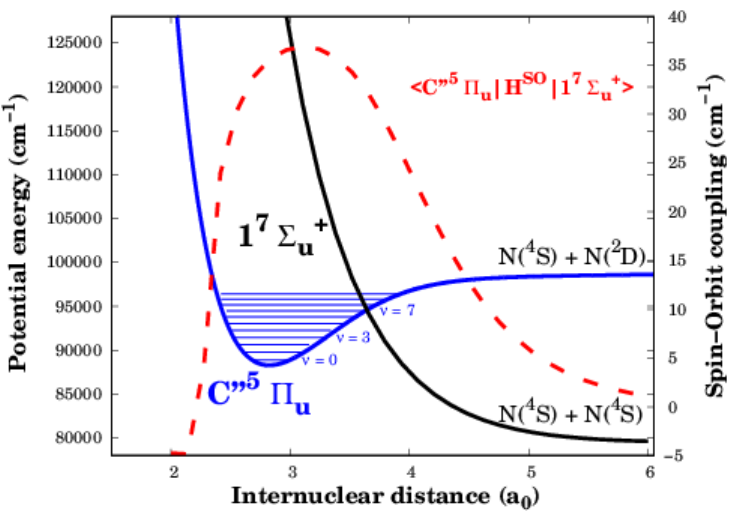}
	\caption{The figure shows the potential energy curves of the electronic states C$''^5\Pi_u$ and 1$^7\Sigma^+_u$ (solid lines). Also shown is the spin-orbit interaction curve between the states (dashed line) as a function of the internuclear distance.}
	\label{FIG:2}
\end{figure}

\begin{figure}
\label{fig:3}
	\centering
	\includegraphics[width=.9\textwidth]{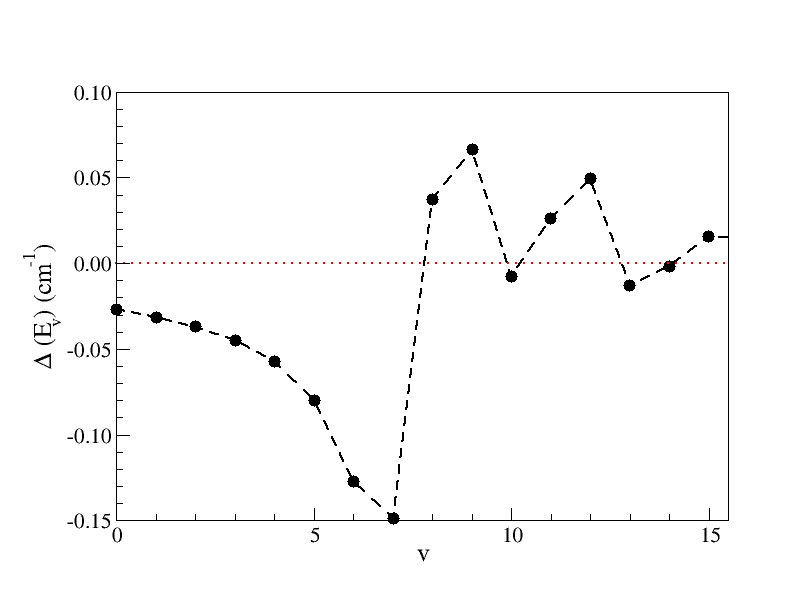}
	\caption{Energy shift in the lowest vibrational levels of the C$''^5\Pi _u$  electronic state due to interaction with the 1$^7\Sigma^+ _u$ electronic state, as a function of the vibrational quantum number.}
	\label{FIG:3}
\end{figure}

\begin{figure}
	\centering
	\includegraphics[width=0.8\textwidth]{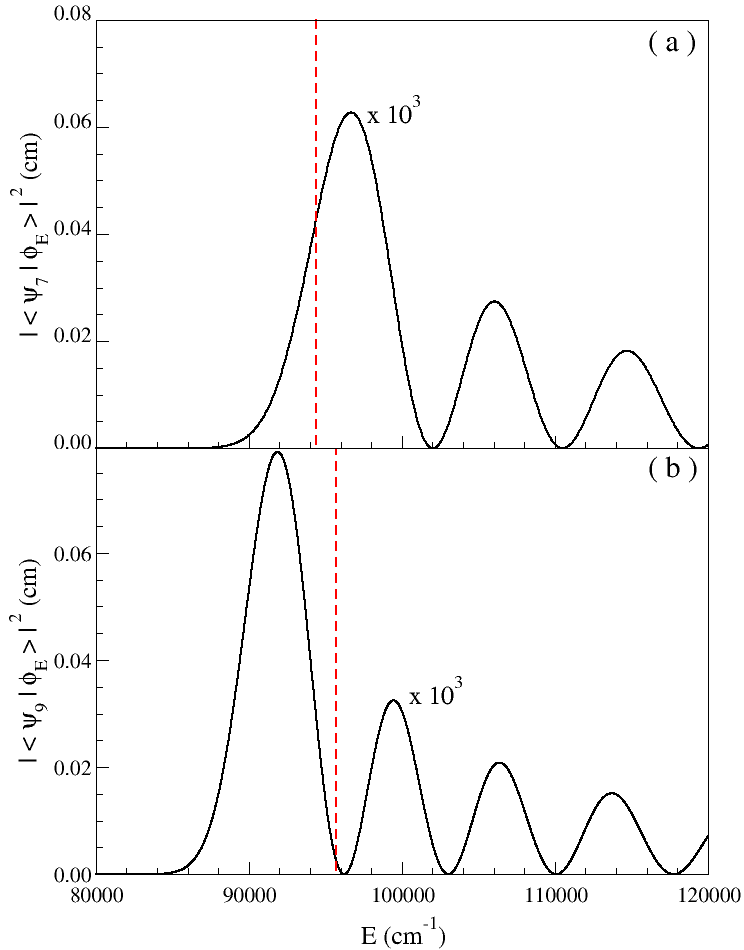}
	\caption{The square of the Franck-Condon factors as a function of the energy for the a) $v$ = 7 vibrational level, b) $v$ = 9 vibrational level. The dashed lines indicates the position of the discrete levels.}
	\label{FIG:4}
\end{figure}

\begin{figure}
	\centering
	\includegraphics[width=1.0\textwidth]{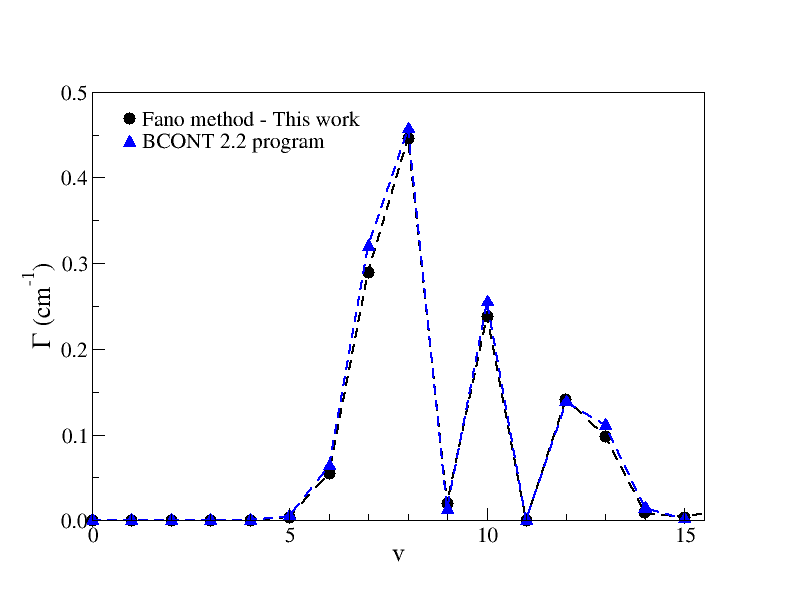}
	\caption{Level width as a function of the vibrational quantum number $v$.}
	\label{FIG:5}
\end{figure}

The most relevant potential energy curves for this work were calculated by Hochlaf {\emph{et al.}}~\cite{hochlaf2010quintet} and da Silva {\emph{et al.}}~\cite{da2020novel}. In these two works, potential energy curves were obtained for the triplet, quintet and septet states, being Hochlaf {\emph{et al.}~\cite{hochlaf2010quintet} the first to present theoretical calculations of the spin-orbit interaction between triplet and quintet states. In a latter study, new calculations were performed by Ventura {\emph{et al.}} \cite{VENTURA2025125986} with new values for the spin-orbit coupling for the C$''^5\Pi _u$ and 1$^7\Sigma ^+_u$ electronic states.

We can now turn our attention to the  energy shifts $\Delta(E_v)$ of the bound vibrational levels $E_v$ of C$''^5\Pi_u$ due to the interaction with the repulsive state 1$^7\Sigma^+_u$ of the N$_2$ molecule. The potential energy  and spin-orbit interaction curves  used in this work for the C$''^5\Pi_u$ and 1$^7\Sigma^+_u$ states were taken from Silva~\textit{et al.}~\cite{da2020novel} and Ventura \textit{et al.}~\cite{VENTURA2025125986}. In summary, these authors reported high-level multireference configuration interaction (MRCI) calculations combined with a larger Dunning basis set (AV5Z). This choice was made because this method has successfully reproduced experimental results with an acceptable degree of reliability. For convenience, these results, up to vibrational level $v$ = 15, are shown in Figure \ref{FIG:2}. It should be noted that Figure \ref{FIG:2} illustrates that the maximum of spin orbit coupling, around 36 cm$^{-1}$, is at internuclear distance where energies curves of he N$_2$ C$''^5\Pi _u$ and 1$^7\Sigma ^+_u$ states intersect to each other. The vibrational energy values $E_v$, obtained as described above, and the corresponding shifts $\Delta(E_v)$ given by Equation (\ref{eq:DeltaE}), are listed in Table \ref{tab:1}.
The latter are also plotted in Figure \ref{FIG:3}.

\section{Predissociation and lifetimes}
\label{predlif}

It can be seen from Table \ref{tab:1} and Figure \ref{FIG:3} that the energy shifts of the electronic state C$''^5\Pi _u$ become more significant starting at the vibrational level $v$ = 5, where the spin-orbit interaction with the dissociative state 1$^7\Sigma^+_u$ reaches the highest values, as shown in Figure \ref{FIG:2}. The maximum value of the energy shift is reached at $v$ = 7. From this vibrational level, the variations in the energy deviation become oscillatory up to the value of $v$ = 14, and then decay monotonically. This behaviour is due to the interaction between the bound levels of the C$''^5\Pi_u$ electronic state and the continuum of the 1$^7\Sigma^+_u$ dissociative state.

A measure of this interaction is provided by the Franck-Condon factor $\bra{\phi_E}\ket{\psi_v}$, which is closely related to the interaction matrix element $V_E^v$. Figure \ref{FIG:4} displays the values of $|\bra{\phi_E}\ket{\psi_v}|^2$ as a function of energy for the vibrational quantum numbers $v$ = 7 and $v$ = 9. In order to provide a sense of the positions of the discrete levels, a dashed line is included in the graphs for this purpose. These quantities exhibit rapid oscillations strongly correlated with the energy. The widths of the vibrational levels, given by $\Gamma = 2\pi|V_{E'}^v|^2$\cite{fano1961effects}, are shown in Figure \ref{FIG:5} as a function of the vibrational quantum number.

In order to analyse the interaction between the two electronic states in greater depth, an alternative procedure was adopted in which the lifetimes were directly computed. As a first step, the lifetimes of the vibrational levels of the electronic state C$''^5\Pi_u$ were calculated from the C$''^5\Pi_u \to$ A$'^5\Sigma^+_g$ vibronic transition employing 
the LEVEL 16 program
\cite{le2017level}. 
The potential energy curves of these electronic states and the transition dipole moment reported by da Silva {\emph{et al.}}~\cite{da2020novel} were used in the LEVEL 16 input file 
for the calculations. The results obtained for the radiative lifetimes ($\tau_{rad}$) for the lowest sixteen vibrational levels of the C$''^5\Pi_u$ ($v$ = 0 - 15) electronic state are presented in the second column ($\tau_{rad}$) of Table \ref{tab:2}.  

These results reveal that the radiative lifetime values remain around a few microseconds up to vibrational levels $v$ = 4. Beyond this quantum number, the radiative lifetime values begin to increase, reaching up to a few milliseconds at $v$ = 10 and and attaining values of a few seconds for $v$ = 15. This behavior is in line with the values predicted in previous work, such as da Silva {\emph{et al.}}~\cite{da2020novel} and Ventura {\emph{et al.}}~\cite{VENTURA2025125986} and reflects the fact that, as the vibrational quantum number of the higher electronic state C$''^5\Pi_u$ increases, the Franck-Condon factors for the transition with the electronic state A$'^5\Sigma^+_u$ become smaller, implying a lower transition probability.

To analyze the pre-dissociation process, the BCONT 2.2 program, developed by 
Le Roy and Kraemer~\cite{le2004bcont}, was employed to calculate the dissociative lifetimes of the C$''^5\Pi_u$ electronic state due to its interaction with the 1$^7\Sigma^+_u$ state. To do this, the potential energy curves calculated by da Silva {\emph{et al.}}~\cite{da2020novel} and the spin-orbit interaction curve reported by Ventura {\emph{et al.}}~\cite{VENTURA2025125986} were used. The results are shown in the third column ($\tau_{pred}$) of Table \ref{tab:2}. The pre-dissociative lifetime values are shown to be extremely high for values of vibrational quantum number $v$ = 2, decreasing to a few microseconds at $v$ = 3 and rapidly being reduced to a few tens of picoseconds for values of $v$ above 6. Based on the values listed in Table \ref{tab:2}, columns 2 and 3, assuming that both the radiative decay process and spin–orbit induced predissociation are active, the effective lifetime $\tau_{eff}$ can be calculated using the following expression:

\begin{equation}
\label{eq:time}
\frac{1}{\tau _{eff}}=\frac{1}{\tau _{rad}}+\frac{1}{\tau _{pred}},
\end{equation}

\noindent with the corresponds results presented in the fourth column of Table \ref{tab:2}.

\section{Discussion}
\label{sec:disc}

The radiative and pre-dissociative lifetimes were obtained using the LEVEL 16 \cite{le2017level} and BCONT 2.2 \cite{le2004bcont} programs, respectively. As discussed above, the energy shifts and the widths of the vibrational levels $\Gamma$  were calculated using the perturbative calculations. The results presented in} Table  \ref{tab:1} and Figure \ref{FIG:3} indicate that the energy shifts become increasingly significant from $v$ = 5, as previously mentioned. Unfortunately, all  high-resolution experimental data \cite{huber1988rotational, pirali2006experimental, pirali2006optogalvanic, pirali2009vibrational,vcermak2018untangling} lack information regarding the vibrational levels of the C$''^5\Pi_u$ state above $v$ = 4, for very low rotational quantum number values $J$, even for the experiments that were not carried out in a supersonic jet~\cite{vcermak2018untangling}. Consequently, there is still no experimental data to corroborate these deviations.

Let us focus our attention on the $\Gamma$ values calculated using the perturbative methods described previously and exhibited in Figure \ref{FIG:5}. One notes that the values for the broadening of the vibrational levels start to deviate from zero at $v$ = 5, where the energy deviations become more pronounced. From then on, the $\Gamma$ values increase, reaching a maximum at $v$ = 8, and subsequently decrease to a small value at $v$ = 9, increasing again at $v$ = 10. It should be emphasized that $\Gamma$ is linked to the average lifetime of the vibrational level through $\hbar/\Gamma$ \cite{fano1961effects}, as well as with the predissociation rate $k_ s$ calculated directly through BCONT 2.2, by the relation $\Gamma = k_s/2 \pi c$, where $c$ is the speed of light.

Moreover, another noteworthy observation can be drawn by examining the fourth column of Table \ref{tab:2}. The effective lifetimes ($\tau_{eff}$), calculated using Equation \ref{eq:time} and the results obtained with the programs LEVEL 16\cite{le2017level} and BCONT 2.2\cite{le2004bcont}, lead to a behavior identical to that of the inverse of $\Gamma$, i.e., the lifetime values decrease from $v$ = 5, reach a minimum at $v$ = 8, increase at $v$= 9 and return, at $v=10$, to a value close to that at $v$ = 5. In order to more effectively compare the results obtained in this work with the calculations from BCONT 2.2, we have added to Figure 5 the values calculated from the relation between the dissociation rates and $\Gamma$, showing excellent agreement among the results. In other words, this behavior is consistently observed through two distinct methods.  

This point can be understood by carefully examining Figure \ref{FIG:4}. Panel (a) shows that the squared Franck-Condon factor reaches a value near its maximum when the continuum energy equals the discrete energy level $E_v$ for $v=7$. This behavior is not observed in panel (b) of this Figure, which presents a similar plot for $v=9$. This vibrational level intersects the squared Franck-Condon curve at values close to zero, indicating a weak interaction. This explains the increase in the effective lifetime ($\tau_{eff}$) as well as the reduction in vibrational level broadening ($\Gamma$), despite having a non-zero energy deviation $\Delta E_v$, as shown in Figure \ref{FIG:3}.

\section{Conclusion}
\label{sec:conc}

In this article the treatment of interference between a bound state and a dissociative state based on the perturbative approach proposed by Fano \cite{fano1961effects} is presented. The method was applied to a problem reported by Child and Lefebvre \cite{child1977shifts}, showing excellent agreement. Subsequently, this procedure was applied to the case of the interaction between C$''^5\Pi_u$ and 1$^7\Sigma^+_u$ eletronic states, where the potential energy curves of both states had already been calculated, as well as the spin-orbit coupling curve between them. In this way, it was possible to calculate the energy shifts $\Delta(E_v)$ for the bound energy levels, the squared Franck-Condon factors $|\bra{\phi_E}\ket{\psi_v}|^2$ of the interaction between the electronic states as well as the  broadening $\Gamma$ of the vibrational levels.

At the same time, calculations of the radiative lifetimes of the transition between the electronic states C$''^5\Pi_u \to$ A$'^5\Sigma^+_g$ and the pre-dissociative lifetimes of the C$''^5\Pi_u$ electronic state, due to its interaction with the dissociative electronic state 1$^7\Sigma^+_u$, were carried out using the LEVEL 16\cite{le2017level} and BCONT 
2.2\cite{le2004bcont} programs, respectively.  These calculations enabled the determination of dissociation rates and the effective lifetimes of the vibrational levels of the C$''^5\Pi_u$ electronic state.
The values obtained fully agree with the estimates derived from the calculation of the $\Gamma$ broadening obtained using perturbative calculations. 

Unfortunately, the absence of experimental data prevents a direct comparison with the results obtained in this work. We believe that further experimental efforts aimed at obtaining more information about the pre-dissociation processes of the C$''^5\Pi_u$ state are necessary.

\newpage

\begin{acknowledgement}

The authors thank the Pr\'o-Reitoria de P\'os-Gradua\c{c}\~ao, Pesquisa e Inova\c{c}\~ao of Universidade Federal Fluminense (PROPPI-UFF), the Conselho Nacional de Desenvolvimento Cient\'{\i}fico e Tecnol\'ogico (CNPq), the Coordena\c{c}\~ao de Aperfei\c coamento de Pessoal de N\'{\i}vel Superior (CAPES) - Finance Code 001, and Fundação de Amparo à Pesquisa do Estado do Rio de Janeiro (FAPERJ) for financial support (Grants number E-26/210.475/2024 and E-26/210.402/2024). R.S. da Silva would like to thank the Fundação de Amparo à Pesquisa do Estado do Rio de Janeiro (FAPERJ, E-26/205.639/2022) for support.  L.R. Ventura is in debt with FAPESP for a post-doctoral fellowship (grant number 2023/08074-5). In special, J. Amorim would like to thank CNPq (Grant number 308548/2023-0), FAPESP (Grant number 2023/08074-5) and PROPPI\-/UFF (Grant 05/2022) for financial support. S. R. Souza thank the Instituto Nacional de Ciência e Tecnologia Física Nuclear e Aplicações (INCT-FNA), Proc. No.464898/2014-5.

\end{acknowledgement}

\newpage

\bibliography{energy_shifts}

\newpage

\begin{table}
\caption{Energy levels given by Eq.\ (\ref{eq:Ev}) and corresponding shifts calculated through Eq.\ (\ref{eq:DeltaE})} 

\centering
\begin{tabular}{ccc|ccc}
\hline 
\label{tab:1}
 $v$ & $E_v$ (cm$^{-1}$) & $\Delta (E_v)$ (cm$^{-1}$) &
 $v$ & $E_v$ (cm$^{-1}$) & $\Delta (E_v)$ (cm$^{-1}$) \tabularnewline
\hline

0 & 453.87 & -0.0267 & 8 & 6757.08 & 0.0372 \tabularnewline

1 & 1342.37 & -0.0312 & 9 & 7390.77 & 0.0663  \tabularnewline

2 & 2204.67 &  -0.0370 & 10 & 7979.18 & -0.0075 \tabularnewline

3 & 3040.02 & -0.0450 &  11 & 8516.90 & 0.0263 \tabularnewline

4 & 3847.27 & -0.0571 & 12 & 8997.28 & 0.0494  \tabularnewline
 
5 & 4624.85 & -0.0800 & 13 & 9412.07 &  -0.0127 \tabularnewline
 
6 & 5370.77 & -0.1270 & 14 & 9751.28 &  -0.0019 \tabularnewline
 
7 & 6082.52 & -0.1490 & 15 & 10005.00 &  0.0156 \tabularnewline

\hline 
\end{tabular} 
\end{table}

\begin{table}[H]

\caption{Radiative ($\tau _{rad}$) and predissociated ($\tau_{pred}$) lifetimes for the vibrational levels of the C$''^5\Pi _u$ calculated as explained in the text and effective lifetimes ($\tau _{eff}$) calculated by using Equation (\ref{eq:time}). All values in $ns$.}

\centering
\begin{tabular}{cccc}
\hline
\label{tab:2}
 $v$& $\tau_{rad}(ns)$ & $\tau_{pred}(ns)$& $\tau_{eff}$(ns)\tabularnewline
\hline

0 & 3172.67& 7.0693$\times10^{10}$ & 3172.67  \tabularnewline

1 &3744.95 &6.8216$\times10^{7}$ &3744.74 \tabularnewline

2& 4829.92 & 2.1586$\times10^{5}$ &4724.22 \tabularnewline
 
3 &4813.47 &1.6765$\times10^{3}$ & 1243.41
 \tabularnewline

4 &5994.13 &28.612 &28.48 \tabularnewline
 
5 &6352.97 &1.0432 &1.043 \tabularnewline

6 &9364.68 &0.0837 &0.084\tabularnewline
 
7 &10085.65 &0.01661 &0.0166 \tabularnewline

8 &23459.13 &0.01162 &0.0116 \tabularnewline

9 & 153.13$\times10^{3}$ & 0.41879 & 0.4188 \tabularnewline

10 & 1.85$\times10^{6}$ & 0.0208 & 0.0208 \tabularnewline

11 & 4.36$\times10^{6}$ & 27.389 & 27.389 \tabularnewline

12 & 5.42$\times10^{6}$ & 0.0384 & 0.0384 \tabularnewline

13 & 1.04$\times10^{8}$ & 0.0480 & 0.048 \tabularnewline

14 & 1.48$\times10^{10}$ & 0.3783 & 0.3783 \tabularnewline

15 & 8.56$\times10^{10}$ & 2.9141 & 2.9141 \tabularnewline
\hline
\end{tabular} \\
\end{table}

\newpage

\newpage

\begin{tocentry}
\centering
\includegraphics[width=0.92\textwidth]{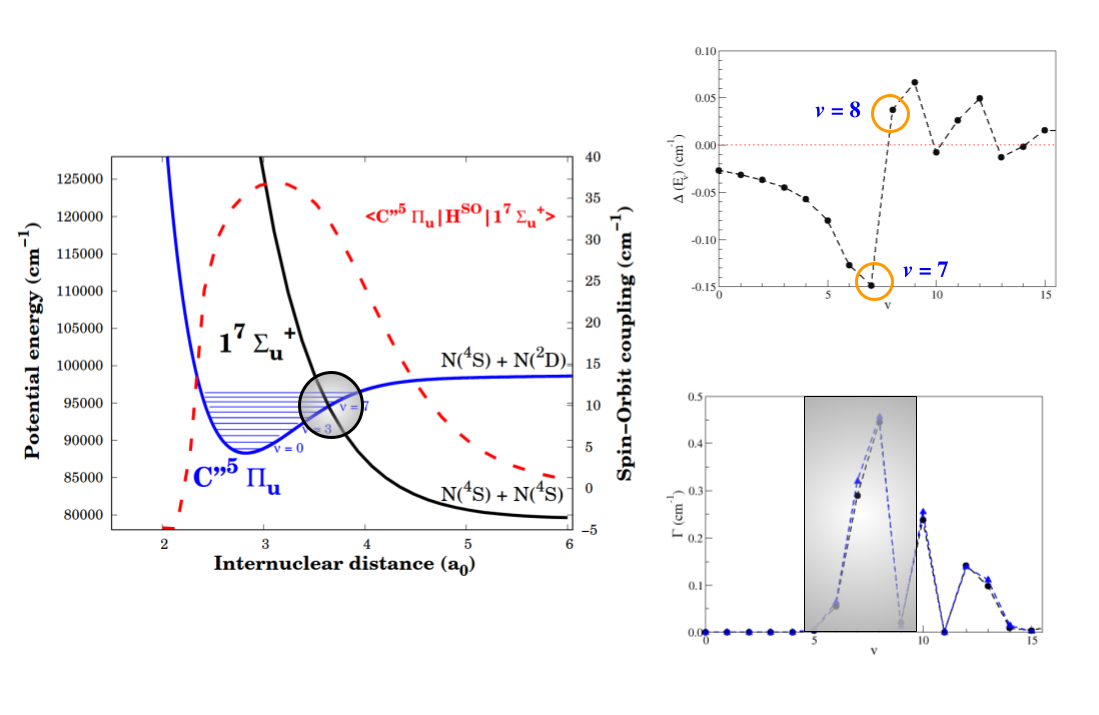}

\end{tocentry}

\end{document}